\documentclass[aps,pra]{revtex4}
\usepackage{graphicx}

\newcommand{\sign}{\mathop{\rm sign}\nolimits}

\begin{document}
%PACS 05.30.Jp 
\title{Super-Tonks-Girardeau regime in trapped one-dimensional dipolar gases}

\author{G.E. Astrakharchik}
\affiliation{Departament de F\'{\i}sica i Enginyeria Nuclear, Campus Nord B4-B5, Universitat Polit\`ecnica de Catalunya, E-08034 Barcelona, Spain}
\author{Yu.E.~Lozovik}%\email{lozovik@isan.troitsk.ru}
\affiliation{Institute of Spectroscopy, 142190 Troitsk, Moscow region, Russia}
\date{\today}

\begin{abstract}
Possible signatures of a super-Tonks-Girardeau gas in bosonic systems of trapped
quasi-one-dimensional dipoles are discussed at zero temperature. We provide  estimation 
of the frequency of the lowest compressional mode and compare it to analytical results derived 
using harmonic approach in the high density regime. We construct an exact mapping of
the ground-state wave function of one-dimensional dipolar system of bosons, fermions
and Bose-Fermi mixture and conclude that local properties and energy are the same at
zero temperature. A question of to which extent the dipolar potential can be treated
long- or short- range is discussed.
\end{abstract}
\maketitle

\section{Introduction}

Recent realization of chromium condensates\cite{Griesmaier05} %in the group in Stuttgart 
have brought much interest to unusual properties of dipolar condensates.
Use of chromium have proved to be very advantageous as it has very large permanent
dipolar moment of $\mu=6\mu_B$. Other possible realizations of dipolar gases includ
polar molecules\cite{Ðaimberger04,Sage05,Buchler07} and laser-induced
dipoles\cite{Tong04,Lukin01}. Study of collective oscillations is a very sensitive
tool for the investigation of the cold gas properties\cite{Giovanazzi07}. Dipolar
forces compete with short-range $s$-wave scattering interaction. Although in the
first experiments the $s$-wave interaction was giving dominant contribution to the
energy, the effects of dipolar interactions have been detected in the shape of
expanding cloud\cite{Stuhler05}. During the expansion of an initially trapped gas
dipolar forces might lead to an anisotropic shape of an expanding
cloud\cite{Yi2003}. Use of Feshbach resonance technique have brought experiment to a
new level. Indeed, by the means of Feshbach resonance technique it is possible to reduce $s$-wave
isotropic contact interaction, such that the anisotropic magnetic dipole-dipole
interaction between $^{52}$Cr atoms becomes comparable in strength\cite{Lahaye07}.
This induces large changes of the aspect ratio of the cloud, and, for strong dipolar
interaction, the inversion of ellipticity during expansion can even be
suppressed\cite{Lahaye07}. 
%By tuning the $s$-wave scattering length to zero one realizes essentially a pure system of dipoles. 
Chromium atoms have have as well fermionic $^{53}$Cr isotope and Bose - Fermi cold mixtures
have been recently realized \cite{Chicireanu06}.

Furthermore, dipolar interactions are very interesting as they contain both repulsive and
attractive parts. Having attractive interactions between bosonic particles at
temperatures so low that condensate is formed might lead to instability and a violent
collapse process\cite{Hulet00,Donley01,Roberts01}. Collapse of large dipolar
condensates in traps have been studied using non-local
Gross-Pitaevskii\cite{Santos00,Lushnikov02,Ronen06,Fischer06} theory and more
precise Diffusion Monte Carlo (DMC) \cite{Ronen06} approach.

One-dimensional (1D) cold systems have received great attention in the last
years~\cite{Bloch05, Moritz03, Richard03}. Role of quantum fluctuations is increased
in reduced dimensionality leading to sometimes very different behavior. For example,
a peculiarity of a one-dimensional world is an absence of a true Bose condensate in
1D homogeneous systems even at zero temperature~\cite{Hohenberg67}. There is a
certain trend to study low-dimensional systems in the last years.

One of the most precise techniques, which can be used for testing the equation of
state, is the measurement of the frequency of the ``breathing'' mode produced by a
sudden change of the frequency of the harmonic trapping. We note that differences of
several percent can be resolved in present high-precision experiments with cold
gases (see, for example, Ref.~\cite{Altmeyer06}).

In this paper we address properties of a quasi-one-dimensional dipolar system at
zero temperature. We provide an explicit mapping of the fermionic ground-state wave
function and a wave function of Bose-Fermi mixture to bosonic wave function in
one-dimension. 
Due to this mapping energy of dipolar systems containing fermions is predicted using results
previously obtained for bosons.
We calculate
frequency  of the lowest breathing mode in the trap and compare it to analytic
predictions obtained in the high density limit. 
Measurement of the frequencies of collective oscillations
can provide a signature of a
super-Tonks-Girardeau (STG) regime\cite{Astrakharchik04d,Batchelor05} described by very
strong correlations (stronger than in the Tonks-Girardeau gas \cite{Girardeau60}
in which the coupling constant is infinitely large). It is very difficult to reach STG
regime in systems with short-range interactions, while its realization is feasible in dipolar systems.
We address problem of superfluidity in one-dimensional systems and a question up to
which extent dipolar interaction is short- or long- range.

\section{Model}

A system of $N$ dipoles in one-dimensional geometry is described by the following
model Hamiltonian:
\begin{eqnarray}
\hat H =
-\frac{\hbar^2}{2M}\sum\limits_{i=1}^N\frac{\partial^2}{\partial z_i^2}
+\frac{1}{2}\sum\limits_{i=1}^N M\omega_z^2z_i^2
+\frac{C_{dd}}{4\pi}\sum\limits_{i<j}\frac{1}{|z_i-z_j|^3}
\label{H}
\end{eqnarray}
Here we assume that all dipoles are polarized and are oriented perpendicularly to
the one-dimensional line. This stabilizes the system (as only repulsive part of
dipolar interactions is relevant) and avoids collapses due to attraction. Expression
for the coupling constant $C_{dd}$ depends on the nature of the dipolar interaction.
For example, following realizations are possible:

1) Cold bosonic atoms, with induced or static dipole momenta, in a transverse trap
so tight that excitations of the levels of the transverse confinement are not
possible and the system is dynamically one-dimensional. The longitudinal confinement
is described by the frequency $\omega_z$ of the harmonic trapping potential. The
dipoles themselves can be either induced or permanent. In the case of dipoles
induced by an electric field $E$ the coupling has the form $C_{dd} = E^2\alpha^2$,
where $\alpha$ is the static polarizability. For permanent magnetic dipoles aligned
by an external magnetic field one has $C_{dd} = m^2$, where $m$ is the magnetic
dipole moment. We suppose that $s$-wave scattering length is tuned to zero by
applying Feshbach resonance and only dipolar forces are relevant.

2) Spatially indirect excitons in two coupled quantum wires. A quantum wire is a
semiconductor nanostructure where an electron or a hole is allowed to move only in
one direction and excitations of the transverse quantization levels are negligible.
In two parallel quantum wires, one containing only holes, and the other only
electrons, holes and electrons couple forming indirect excitons. If such a system is
dilute enough, it constitutes a $1D$ set of dipoles. In this case $C_{dd}
= e^2d^2/\varepsilon$, where $e$ is an electron's charge, $\varepsilon$ is the
dielectric constant of the semiconductor, and $d$ is the distance between the
centers of the quantum wires. This system is 1D counterpart of 2D indirect exciton
system in coupled quantum wells.
%, which was extensively studied both theoretically \cite{excitonstheory} and experimentally\cite{excitonsexper}.

\section{Super-Tonks-Girardeau regime}

Properties of a homogeneous system ($\omega_z=0$) have been studied numerically
in\cite{Arkhipov05} by means of Diffusion Monte Carlo method. The equation of state
and correlation functions have been calculated as a function of the guiding
parameter $nr_0$, where $r_0 = M C_{dd}/(4\pi\hbar^2)$ is a characteristic length,
$n$ being the linear density. It has been found that the system is extremely
correlated and shows crystal-like properties as the strength of dipolar interactions
$nr_0$ is increased. This behavior is very different from the one of systems with
short-range potentials, like the ones recently realized
experimentally\cite{Tolra04,Kinoshita05}. Indeed, one-dimensional systems with
$\delta$-interaction, $V(z)=-2\hbar^2/ma_{1D}\delta(z)$, (Lieb-Liniger
model\cite{Lieb63}) have a completely different behavior of correlation
functions\cite{Astrakharchik03,Astrakharchik06b} which in this case
is much closer to the one of a
weakly interacting Bose gas. For example a typical shape of the static structure
factor $S(k)$ is a smooth function which goes from zero for $k=0$ to asymptotic
constant value $S(k)=1$ for large $|k|$. In the most strongly interacting limit of
Lieb-Liniger system $-2\hbar^2/ma_{1D}\to+\infty$ (Tonks-Girardeau
regime\cite{Girardeau60}) the static structure factor is equal to the one of an
ideal Fermi gas and has a discontinuity in the derivative at momentum $|k|=2k_F=2\pi n$. Even
stronger correlations might be achieved in a system with short-ranged potential by
quickly crossing the confined induced resonance\cite{Olshanii98} as proposed in
\cite{Astrakharchik04d}. In this resulting ``super-Tonks-Girardeu'' regime
bosonic atoms will interact with attractive potential $-2\hbar^2/ma_{1D}\to-\infty$.
The true ground state in this regime is a soliton-like state with large and negative
energy\cite{McGuire64} (``attractive collapse'' of the system). Instead, if the sweep across resonance is fast and does
not significantly perturb position of the particles, the system will still remain 
in a gas-like state (which is metastable). As shown in Ref.~\cite{Astrakharchik04d}, this state
is dynamically stable if the gas parameter $na_{1D}$ is relatively small. The static
structure factor in super-Tonks-Girardeau regime has a peak at $|k|=2k_F$ and height
of the peak increases for larger values of the gas parameter $na_{1D}$. The
super-Tonks-Girardeau regime also exists in a number one-dimensional system (bosonic
and fermionic) with infinite strength potentials, namely hard-rods\cite{Mazzanti07}
and Calogero-Sutherland model\cite{Astrakharchik06c}. So far this regime have never
been observed experimentally with the best candidate for its observation being a
system of dipoles. An experimental signature of STG regime in a trapped system is a
frequency of lowest compressional mode $\Omega_z$ larger than
$2\omega_z$\cite{Astrakharchik04d}. In the Section~\ref{osc} we calculate explicitly
the dependence of $\Omega_z$ on parameters of a trap.

Super-Tonks-Girardeau regime is expected to have quite unusual properties. It has
been shown\cite{Neto94,Astrakharchik02c,Cazalilla04} for a Lieb-Liniger model that
the dynamic form factor has a power low in the point where the excitation spectrum
touches zero $S(\omega,2k_F)\propto\omega^{\eta-2},\omega\to 0$, where $\eta=2\hbar
k_F/mc$. In the regime of repulsive $\delta$-interaction $\eta>2$ and $S(0,2k_F)=0$.
For the marginal case of Tonks-Girardeau regime $S(\omega,2k_F)=const$. Instead for
attractive $\delta$-interaction ({\it i.e.} in the Super-Tonks-Girardeau regime) this
expression predicts a weak (power law) divergence in the dynamic form
factor $S(\omega,2k_F)\to\infty, \omega\to 0$. For additional information on
singularities in the dynamic structure factor for different values of momenta
see Ref.~\cite{Pustilnik06,Khodas07}

The problem of superfluidity has peculiarities in one-dimensional system.
While different ways to define superfluid part in three- and two- dimensional
systems are consistent, this is not the case in a one-dimensional world. Indeed,
its calculation as a response of a liquid to sample boundary motion ({\it
winding-number} method\cite{Pollock87}) for the highly correlated states described
by translationary invariant gas-like wave function (for example, exact wave function
for the Tonks-Girardeau\cite{Girardeau60},
Calogero-Sutherland\cite{Calogero69,Sutherland71}, hard-rods\cite{Girardeau60} systems; and
DMC evaluation for Lieb-Liniger\cite{Astrakharchik03}, dipolar
\cite{Arkhipov05} systems) would provide the result that such systems are completely
superfluid. This argument would apply even to the TG regime, where bosonic system
has many similar properties to an ideal Fermi gas. At the same time, one should keep in mind
that exposure of an one-dimensional ideal Fermi gas to a tiny perturbation will
change the ground-state wave function in a dramatic way: the overlap of the new wave
function and the old is essentially zero. This effect known as orthogonality
catastrophe (see, for example, textbook \cite{Levitov02eng}) shows that in one-dimensional 
system stability to external perturbations have to be carefully checked. Contrary
to winding number approach, Landau argument would lead to exactly opposite result
classifying systems as completely normal. Indeed, in one-dimensional Luttinger
liquids the excitation spectrum always touches zero at finite value of momentum,
$|k|=2k_F$ (see, for example, Ref.~\cite{Cazalilla04}; for Lieb-Liniger model the
``type II'' excitation\cite{Lieb63b} that touches zero can be identified as a dark
soliton\cite{Ishikawa80}). Another way to calculate superfluid density by the
response to transverse probe (transverse current-current response) is not applicable
within one-dimensional description as no transverse direction is included in the
model. Contradictions in the results arise from different definitions of superfluid
part and reflect the non-standard nature of the system. Probably the most natural
and appropriate way to test the superfluidity in 1D world is done by dragging a
small impurity (perturbation) through the system and seeing if this leads to energy
dissipation. The force $F_V$, experienced by the system, depends on the interaction
parameter $\eta$ as $F_V\propto V^{\eta-1}$\cite{Astrakharchik02c} where $V$ is
velocity with which a small $\delta$-perturbation moves through the system. Thus in
the mean-field limit $\eta\to\infty$ the force is vanishing and from a practical
point of view system behaves analogously to a superfluid. On the contrary, in the
Tonks-Girardeau limit $F_V \propto V$ and system behaves, from the point of view of
friction, as a normal system, where the drag force is proportional to the velocity.
In between there is a smooth crossover. The dipolar one-dimensional systems
are expected to behave as normal ones.

Although the Bose-Einstein condensation is absent in a one-dimensional system even
at zero temperature, its reminiscence still can be observed in divergence of the momentum
distribution for $k=0$. This divergence is present in a homogeneous system of bosons
in the Tonks-Girardeau regime. From the Luttinger liquid theory it is possible to
show that deep in super-Tonks-Girardeau regime this divergence will be removed. This
happens for $\eta=1$ (similarly to Calogero-Sutherland \cite{Astrakharchik06c} and
Hard-Rod \cite{Mazzanti07} systems) or in terms of Luttinger parameter $K=1/2$.
The Luttinger parameter can be extracted from the equation of state or from the
phononic part of the static structure factor and is explicitly
given in Ref.~\cite{Citro07}.

\section{Frequencies of collective oscillations\label{osc}}

\begin{figure}
\begin{center}
\includegraphics*[width=0.4\columnwidth,angle=0]{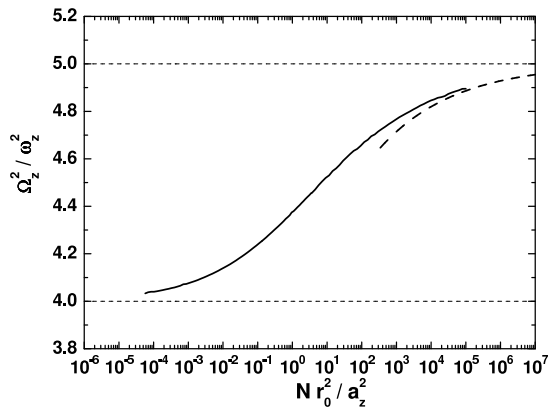}
\caption{Square of the lowest breathing mode frequency, $\Omega^2_z$, as a function of the coupling
strength $Nr_0^2/a_z^2$. The solid line: using equation of state of one-dimensional
dipols (data is taken from \cite{Astrakharchik:candidate}, Fig.~6.5), dashed line:
high density expansion, Eq.~(\ref{freq_pert}).}
\label{Fig1}
\end{center}
\end{figure}

The equation of state obtained in \cite{Arkhipov05} can be used to predict
properties of a trapped one-dimensional gas within local density approximation
(LDA) (see, for example, \cite{Astrakharchik06a}) by assuming that the chemical potential in a trap is a
sum of local chemical potential, taken to be the same as in a homogeneous system
$\mu_{hom}(n)$, and the external harmonic potential: $\mu =
\mu_{hom}(n(z))+(1/2)M\omega_z^2z^2$. Within LDA number of particles $N$ and
oscillator length $a_z=\sqrt{\hbar/M\omega_z}$ come in a single
combination, $Nr_0^2/a_z^2$. The LDA is expected to describe correctly properties of
a gas in the trap if size of the cloud is large compared to $a_z$. 
From the knowledge of the density profile $n(z)$ one can obtain the mean square
radius of the cloud $\langle z^2\rangle=\int_{-R}^R n(z)z^2 dz /N$ and thus, making
use of the result \cite{Menotti02} $\Omega^2_z=-2 \langle z^2\rangle / (d\langle
z^2\rangle/d\omega_z^2)$ to calculate the frequency $\Omega^2_z$ of the lowest
breathing mode.

Fig.~\ref{Fig1} shows the square of the lowest breathing mode frequency $\Omega^2_z$
as a function of the coupling strength $Nr_0^2/a_z^2$. In the TG regime,
$Nr_0^2/a_z^2\ll 1$, the frequency reaches a typical result of an ideal Fermi gas
$\Omega_z=2\omega_z$. The observation of a breathing mode with a frequency larger
than $2\omega_z$ would be a clear signature of the super-Tonks-Girardeau regime.

Attempt to calculate analytically the equation of state in the regime $nr_0\ll 1$ in terms
of the $s$-wave scattering length and to obtain expansion for the frequencies 
of collective oscillations encounters problems as will be explained in Section~\ref{secLR}. Instead some
analytical results can be obtained in the regime of high densities $nr_0\gg 1$. In
this regime the dipolar system has crystal-like properties\cite{Arkhipov05}. At the same time the
difference in the energy between gas-like and crystal-like descriptions is extremely
small. This justifies an attempt to derive 
in a perturbative way 
the equation of state of a crystal 
and use it for an approximate description of the
gas-like phase at the same density.
We use classical crystal harmonic approach to calculate equation of state in the
perturbative manner.
Details of the calculation are provided in Appendix~\ref{secAppendix}. The result
for the equation of state is
\begin{eqnarray}
\frac{E}{N} = \zeta(3)\frac{\hbar^2}{Mr_0^2}(nr_0)^3 + C \frac{\hbar^2}{Mr_0^2}
(nr_0)^{5/2}+..., nr_0\gg 1,
\label{pert}
\end{eqnarray}
with $C=2.26...$. 
Application of the LDA to ``perturbative'' equations of
state has been studied in details in \cite{Astrakharchik06a}. Density profile, total
and release energy are easily obtained by using expansion (\ref{pert}) in the high
density $Nr_0^2/a_z^2\gg 1$ regime. Here we report the frequency of the lowest
compressional mode:
\begin{eqnarray}
\frac{\Omega^2_z}{\omega_z^2}
= 5 - \frac{1.1358...}{(Nr_0^2/a_z^2)^{1/5}}+...
\label{freq_pert}
\end{eqnarray}
The obtained dependence is shown in Fig.~\ref{Fig1} as a dashed line and provides good
description for large densities.

\section{Bose-Fermi mapping for the ground-state wave function}

In the previous paper\cite{Arkhipov05} we have calculated zero-temperature equation
of state for the homogeneous Hamiltonian~(\ref{H}) with $\omega_z=0$ for bosonic
particles and have sampled correlation functions over the ground-state wave function
$\Psi_B(z_1,...,z_N)$. As in experiments dipoles can be not only bosons,
but as well
fermions, or even a mixture of bosons and fermions we note that the equation of
state we have obtained is applicable also to systems containing fermions. In order
to prove that we construct an exact mapping of the wave function of bosonic dipoles to a wave
function of fermionic wave function by analogy to what Marvin Girardeau did in
his classical work\cite{Girardeau60}. For simplicity we start with a system of
same-spin fermions. Such a system is described by the Hamiltonian (\ref{H}). The
main difference from the bosonic case is that fermionic wave function must be
antisymmetric with the respect of exchange of any two particles. This can be done as
\begin{eqnarray}
\Psi_F(z_1,...,z_N)=\prod_{i<j}\sign(z_i-z_j)\Psi_B(z_1,...,z_N).
\label{psi}
\end{eqnarray}
It is easy to check that the symmetry of wave function (\ref{psi}) is correct.
Furthermore, due to Pauli exclusion principle, two Fermions are not permitted to
stay in the same place $\Psi_F(z_1,...,z_N)=0$ if $z_i=z_j$. This is already
satisfied in the construction of bosonic wave function $\Psi_B(z_1,...,z_N)$ due to
divergence of $1/|z|^3$ interaction for small $z$. It means that performed Diffusion
Monte Carlo calculation for bosons is equivalent to Fixed-Node Diffusion Monte Carlo
calculation for fermions with exactly known nodal structure as far as energy and
local quantities (pair correlation function, static structure factor, {\it etc.})
are concerned.  The trick (\ref{psi}) has been successfully used for study Fermionic
Calogero-Sutherland model in Ref.~\cite{Astrakharchik06c}.

Another system which will have similar zero-temperature equation of state is a
system of two-component fermions, where the mass $M_\sigma$ of different spin atoms
$\sigma=\uparrow,\downarrow$ is the same. Hamiltonian of such a system is given by
\begin{eqnarray}
\hat H =
-\frac{\hbar^2}{2M}
\sum\limits_{\sigma; i=1}^{N^\sigma}\frac{\partial^2}{\partial z_{i,\sigma}^2}
+\frac{1}{2}M\omega_z^2
\sum\limits_{\sigma;i=1}^{N^\sigma} z_{i,\sigma}^2
+\frac{C_{dd}}{8\pi}
\left(
\sum\limits_{i\ne j}^{N^\uparrow}\frac{1}{|z_i^\uparrow-z_j^\uparrow|^3}
+\sum\limits_{i\ne j}^{N^\downarrow}\frac{1}{|z_i^\downarrow-z_j^\downarrow|^3}
+\sum\limits_{i,j}\frac{1}{|z_i^\uparrow-z_j^\downarrow|^3}
\right)
\label{Hfermi}
\end{eqnarray}
Ground state wave function of system with totally $N=N^\uparrow+N^\downarrow$ atoms
is than mapped onto a system of $N$ bosons with a Hamiltonian (\ref{H}) as
\begin{eqnarray}
\Psi_F(z_1^\uparrow,...,z_{N^\uparrow}^\uparrow,z_1^\downarrow,...,z_{N^\downarrow}^\downarrow)
=
\prod_{i<j}^{N^\uparrow}\sign(z_i^\uparrow-z_j^\uparrow)
\prod_{i<j}^{N^\downarrow}\sign(z_i^\downarrow-z_j^\downarrow)
\Psi_B(z_1^\uparrow,...,z_N^\uparrow,z_1^\downarrow,...,z_N^\downarrow).
\label{psi2}
\end{eqnarray}

Thus we conclude that a system one-dimensional Fermionic dipoles has the same
ground-state equation of state as a system of bosonic dipoles. Consequently the LDA density
profile in a trap is the same. This means that the frequencies of the lowest
breathing mode for fermions follow the dependence shown in Fig.~\ref{Fig1} with $N$
being total number of dipoles. The limit $Nr_0^2/a_z^2\to 0$ corresponds to ideal
fermions and ideal fermions has the spherical breathing mode $\Omega=2\omega_{ho}$
in any dimension (see, for example, \cite{Astrakharchik06a}). For a finite value of
$Nr_0^2/a_z^2$ the frequency is increased due to repulsive interactions. Similar
effects have been predicted in Ref.~\cite{Astrakharchik04b} using Bethe-{\it ansatz}
theory for systems of two-component fermions with $\delta$-pseudopotential
attractive\cite{Gaudin67,Krivnov75} and repulsive \cite{Yang67} interaction between
atoms of different spin. While homogeneous system of bosons collapses if the
interaction is attractive, this is not the case for fermions, where Pauli principle
stabilizes the system. Adding weak attraction between atoms leads to softening the
breathing mode $\Omega_z<2\omega_z$\cite{Astrakharchik04b}.

The Bethe {\it ansatz} method permits to find ground-state energy and, thus, study
collective oscillations within LDA in a mixture of one-dimensional bosons and
fermions with $\delta$-pseudopotential repulsive interactions and arbitrary
bosons-fermions density ratio\cite{Imambekov06}. Such a mixture was found to be
always stable against demixing\cite{Imambekov06}. Ground-state wave function of a
mixture of one-dimensional bosons and fermions can be obtained with the same
reasoning as for (\ref{psi}) and (\ref{psi2}), thus leading to the same frequency of
oscillations as in Fig.~\ref{Fig1}. Such kind of mapping turns out to be quite
general. Indeed, it works for one-dimensional systems where one-dimensional
interaction potential diverges when two particles meet and shows that energy, local
quantities (pair-correlation function, three-particle correlation function, {\it
etc.}) are the same for bosons and fermions at zero temperature.

\section{Are dipolar interactions long-range?\label{secLR}}

It is common to oppose long-range dipolar interactions to short-range interactions
described by $s$-wave scattering length. But are dipolar interactions really long
range? Or to which extend are they long-range? There are at least two ways to approach this
question.

The first way is to classify potential $V_{int}(r)$ as long- or short- range
depending if the chemical potential is extensive on intensive quantity. If the
homogeneous properties of a large system can be defined by density $n$ only 
(in appropriate units), $\mu = \mu(n)$, the potential is short-range. If instead number of particles $N$ have to be
explicitly specified, 
$\mu= \mu(N,n)$, 
due to strong (diverging) dependence on $N$,  
the potential is of a long-range. This can be
immediately checked by testing the convergence of the potential energy at large
distances:
\begin{eqnarray}
I=\int\limits_{L_{min}}^\infty V_{int}(r) r^{D-1}\;dr,
\label{LR}
\end{eqnarray}
where $L_{min}$ is some cut-off length and $D$ is dimensionality. Following this
definition potential is short-range if it decays at large distances faster than
$r^{-D}$ in $D$ dimensions (see, for example, \cite{Ruffo02}). From this point of view, $1/|r|^3$
potential is long-range in $3D$, while it is short-range in $1D$ and $2D$.

Alternatively, 
short-range potentials can be defined as potentials that can be
described by the asymptotic phase shift. This means that a short-range potential of
range $R$ can be approximated at large distances $r\gg R$ by a free-wave with an
appropriate phase shift, or, being the same, the $s$-wave scattering length $a$. At
sufficiently small densities $na^D \to 0$, the only relevant
length is $a$ and properties (for example energy) can be expressed in terms of the
gas parameter $na^D$. We use definition of the $s$-wave scattering length as a
position of the node of analytic continuation of the scattering solution from
distances larger than the range of the potential in the zero-energy scattering
limit. This definition works well in three-dimensional systems, but also it is
applicable to low-dimensional systems.

It is possible to solve the two-body scattering problem for $1/|r|^3$ potential at
zero energy and find the scattering solution $f(r)$. One needs to look for a regular
solution ($f(0)=0$) of the following differential equation:
\begin{eqnarray}
%\left\{
%\begin{array}{ccc}
%-f''(r)-\frac{D-1}{r}f'(r)+f(r)/r^3&=&0\\
%f(0)&=&0
%\end{array}
%\right.
-\frac{\partial^2f(r)}{\partial r^2}-\frac{D-1}{r}\frac{\partial f(r)}{\partial r}+\frac{f(r)}{r^3}=0,
\label{scat}
\end{eqnarray}
where length is expressed in units of $r_0$ and energy in $\hbar^2/Mr_0^2$. Solutions of
(\ref{scat}) can be written explicitly
%$f^{3D}(r)\propto K_1(2/\sqrt{r})\sqrt{r}$,
%$f^{2D}(r)\propto K_0(2/\sqrt{r})$,
%$f^{1D}(r)\propto K_1(2/\sqrt{r})/\sqrt{r}$,
$f^{3D}(r)\propto r^{1/2}K_1(2r^{-1/2})$,
$f^{2D}(r)\propto K_0(2r^{-1/2})$,
$f^{1D}(r)\propto r^{-1/2}K_1(2r^{-1/2})$,
where $K_n(r)$ denotes the modified
Bessel function of the second kind. In order to find the $s$-wave scattering length
one has to expand $f(r)$ far from the range of the potential ($r\to\infty$) and
compare it to the similar expansion of a plane-wave in appropriate number of
dimensions: $f_{free}^{3D}(r)\propto 1-a_{3D}/r$, $f_{free}^{2D}(r)\propto
\ln(r/a_{2D})$, $f_{free}^{1D}(r)\propto r-a_{1D}$. Expanding the
solutions of Eq.~(\ref{scat}) we find
$f^{3D}(r)\propto 1+(2\gamma-1)/r-(\ln r)/r+{\cal O}(r^{-2})$,
$f^{2D}(r)\propto    2\gamma     -\ln r    +{\cal O}(r^{-1})$,
$f^{1D}(r)\propto r+(2\gamma-1)  -\ln r    +{\cal O}(r^{-1})$,
where $\gamma=0.577...$ is Euler's constant. There are logarithmic terms appearing
in all dimensions. In a two dimensional system such a term is compatible with
asymptotic behavior of a free-wave solution and it is possible to define a finite
scattering length $a_{2D}=e^{2\gamma}r_0=3.172...r_0$. Indeed, numerical evaluation
of the equation of state in a two-dimensional dipolar system\cite{Astrakharchik07}
is in agreement with equation of state for hard-disks with same values of $a_{2D}$\cite{Pilati05}. Also,
Bogoliubov theory at small $na_{2D}^2$ provides correct predictions for the
correlation functions and condensate fraction. Thus, in two-dimensions dipolar
$1/|r|^3$ potential can be treated as a short-range one. Of course, values of the
typical value of the gas parameter $na_{2D}^2$ for which 
universal in terms of $na_{2D}^2$ description starts
being valid is smaller for dipolar interaction, than for ``usual'' short-range
potentials (for example, a soft-disks\cite{Pilati05}).

Instead, in a one-dimensional system the free-wave solution does not contain
logarithmic term and this makes the dipolar solution be incompatible with it. For example,
for short-range potentials  
$s$-wave scattering length can be calculated through
the limit $\lim\limits_{r\to\infty} [r-f(r)/f'(r)]=a_{1D}$. As it is easy to see,
corresponding expression diverges in the case of dipolar interaction potential. It means that it
is not possible to describe properties of a one-dimensional dipolar system by a
short range potential. The same conclusion can be reached by comparing the
ground-state energy of a repulsive dipolar system \cite{Arkhipov05,Citro07} to
ground-state energy of the repulsive short-range
$\delta$-pseudpotential\cite{Lieb63} in the dilute regime $na_{1D}\ll 1$. The
leading term in the energy $E$ is the one of the Tonks-Girardeau gas
$E_{TG}/N=\pi^2\hbar^2n_{1D}^2/6M$\cite{Girardeau60}. The first correction should include
(if the short-range description is possible) terms $na_{1D}$. Instead it is clear
that it is not the case, as for Lieb-Liniger gas $a_{1D}<0$ and energy is lowered
$E<E_{TG}$. By simply neglecting the logarithmic term for dipoles one would find
negative scattering length $a_{1D}=(1-2\gamma)r_0=-0.154...$, and the energy would
be lower. But we definitely know that this is not the case, instead energy is higher
$E>E_{TG}$. Thus we find problems trying to describe the dipolar interaction
potential by a short-range model. Most probably, the same will happen in a three
dimensional system.

\section{Conclusions}

In conclusions, we have studied possible signatures of a super-Tonks-Girardeau gas
in a system of trapped quasi-one-dimensional dipoles at zero temperature. This
regime can be entered by exploiting a confinement induced resonance of the effective
1D scattering amplitude. Using previously calculated equation of state we provide
predictions for the frequency of the lowest compressional mode. Properties
in the high density regime are calculated within harmonic approximation. We provide
an explicit mapping of the ground-state wave function of one-dimensional dipolar
system of bosons; fermions and Bose-Fermi mixture and conclude that local properties
and energy are the same at zero temperature. A question to which extent the dipolar
potential can be treated long- or short- range is discussed. Superfluidity in
one-dimensional systems is tested using different definitions.

During preparation of the paper to publication a related article\cite{Pedri07}
appeared, with a study of frequencies of collective oscillations. Their findings for
the lowest breathing model are in agreement with ours. In
addition frequencies of higher modes are calculated.

%cknowledgments: We thank I.L.Kurbakov for useful discussions.
%We thank B.~Jackson for reading the manuscript.

The work was partially supported by (Spain) Grant No. FIS2005-04181, Generalitat de
Catalunya Grant No. 2005SGR-00779 and RFBR. G.E.A. acknowledges post doctoral
fellowship by MEC (Spain).

\appendix
\section{Equation of state in the high-density regime\label{secAppendix}}

In the high density regime, $nr_0\gg 1$, potential energy dominates and properties of
the system can be compared to the ones of a classical crystal with the lattice
spacing $a$ defined by the density $a=n^{-1}$. Several terms of the expansion of the
equation of state can be found using harmonic approximation (for 2D dipolar crystal
this has been done in Ref.~\cite{Mora07}). The leading term is given by 
potential energy of a classical lattice
\begin{eqnarray}
\frac{E^{(0)}}{N} = \frac{\hbar^2r_0}{M}\sum\limits_{j=1}^\infty\frac{1}{(j a)^3}
=\zeta(3)\frac{\hbar^2}{Mr_0^2}(nr_0)^3
\label{E0}
\end{eqnarray}

\begin{figure}
\begin{center}
\includegraphics*[width=0.4\columnwidth,angle=0]{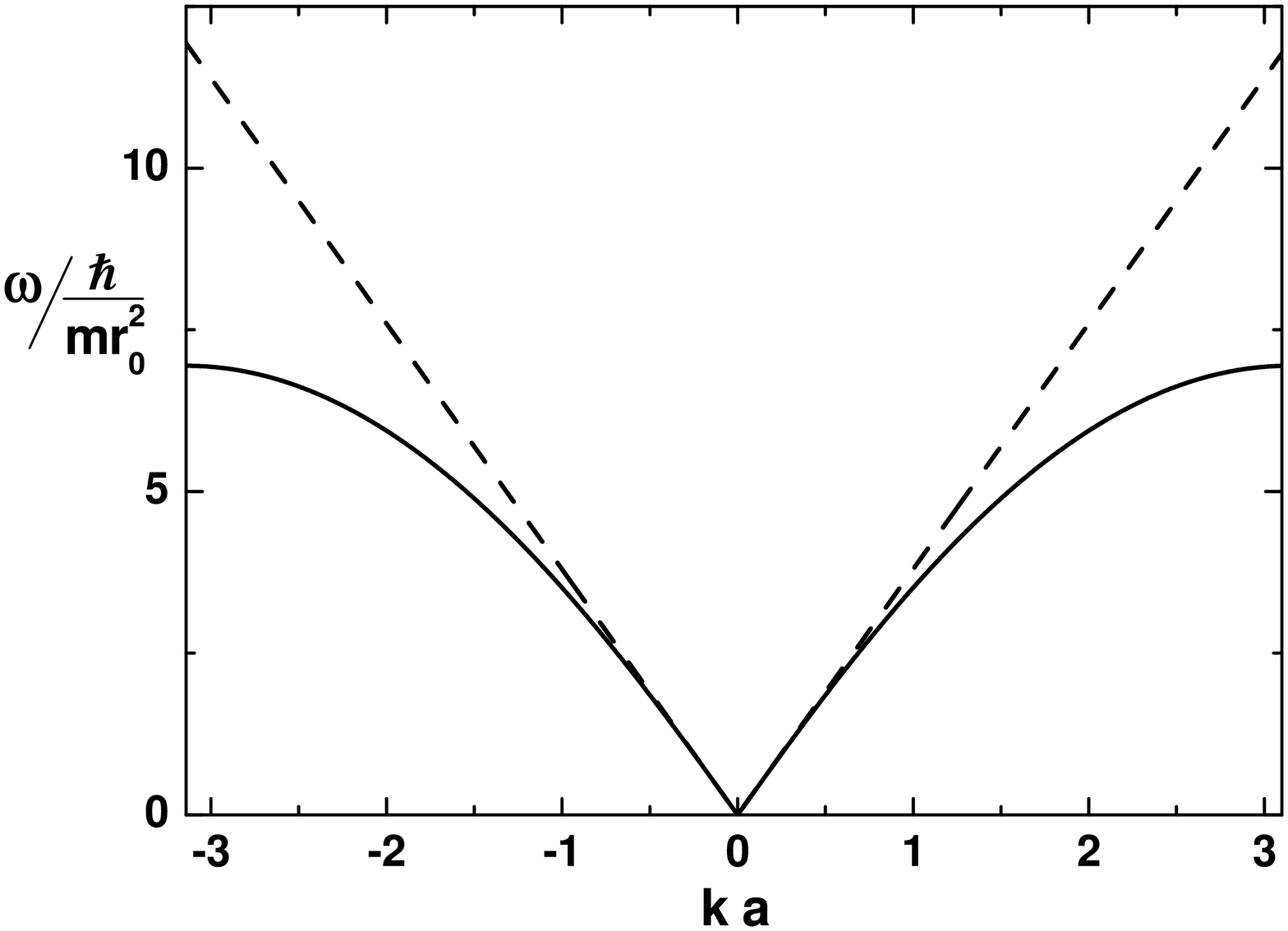}
\caption{Dispersion curve $\omega(k)$ in the first Brillouin zone
$-\frac{\pi}{a}<k<\frac{\pi}{a}$ in the high density limit $nr_0\gg 1$. Solid line:
harmonic approximation, Eq.~(\ref{harmonic}), dashed line: phonons,
Eq.~(\ref{phonons}).}
\label{Fig2}
\end{center}
\end{figure}

Particles move close to lattice sites $u_j = z_j - ja\ll a$. The potential energy
can be expanded up to quadratic terms in $u_j$ and classical equation of motion can
be solved looking for the wave solution $u_j = r_0 e^{i(kja-\omega t)}$. For the
dipolar potential this leads to the following expression for the frequency $\omega$:
\begin{eqnarray}
\omega^2(k) =
\frac{12\hbar^2r_0}{M^2a^5}
\left(
\sum\limits_{j=1}^\infty \frac{2}{j^5}
-\sum\limits_{j=1}^\infty \frac{e^{j(ika)}}{j^5}
-\sum\limits_{j=1}^\infty \frac{e^{j(-ika)}}{j^5}
\right)
=
\frac{12\hbar^2r_0}{M^2a^5} (2\zeta(5) -Li_5(e^{ika}) -Li_5(e^{-ika})),
\label{harmonic}
\end{eqnarray}
where $\zeta(z)$ is the Riemann zeta function and $Li_n(z)$ is a polylogarithm
function. The dispersion relation (\ref{harmonic}) is shown in Fig.~\ref{Fig2}.

The low-momenta behavior $|k|\to 0$ can obtained by expanding exponents in the sums
in~\ref{harmonic} and noticing that zero- and first- order terms get canceled
\begin{eqnarray}
\omega(k) \approx
\sqrt{\frac{12\hbar^2r_0}{M^2a^5} \sum\limits_{j=1}^\infty\frac{k^2a^2}{j^3}}
= 2\sqrt{3\zeta(3)}\frac{\hbar}{Mr_0}(nr_0)^{3/2}|k|
\label{phonons}
\end{eqnarray}
Small $|k|$ behavior correspond to phonons. Same result can be recovered by calculating the
compressibility in the system $mc^2=n \partial \mu / \partial n$, using the leading
term in the energy (\ref{E0}) to calculate the chemical potential $\mu = \partial
E^{(0)} / \partial N$. Then the phononic spectra calculated as $\omega(k)=c|k|$
exactly coincides with expression (\ref{phonons}). In Fig.~\ref{Fig2} the phonon
excitation spectrum is plotted against the solution (\ref{harmonic}). The latter
solution has a characteristic rounding close to boundary of the Brillouin zone. This
feature, of course, is missed in phononic description which becomes inapplicable for
such large values of $|k|$.

Contribution $E^{(1)}$ to the lattice energy (\ref{E0}) is obtained by summation of
the energy of the zero-point motion of atoms. Thus, one has to integrate the dispersion
(\ref{harmonic}) over the first Brillouin zone (BZ):
\begin{eqnarray}
\frac{E^{(1)}}{N}
= \int\limits_{BZ}\frac{\hbar\omega(k)}{2}\frac{dk}{V_{BZ}}
= C \frac{\hbar^2}{Mr_0^2} (nr_0)^{5/2}
\label{E1}
\end{eqnarray}
The obtained contribution is positive, as it adds positive kinetic energy and
describes displacement of atoms from the minimum of potential energy. Correction
(\ref{E1}) scales with density as $(nr_0)^{5/2}$ compared to the $(nr_0)^{3}$
dependence of the dominant term. The coefficient of proportionality $C=2.26...$ was
obtained by numerical integration of (\ref{E1}) with the dispersion relation (\ref{harmonic}).

%\bibliography{astra}

\end{document}